# X-ray Coherent Attosecond Pulse Pair Spectroscopy

Zain Abhari[1*], Thomas M. Linker[2], Thomas Kroll[3], Yurina Michine[4], Gota Yamaguchi[5], Yuichi Inubushi[5], Taito Osaka[5], Jumpei Yamada[6], Ichiro Inoue[5,7], Makina Yabashi[5,8], Aliaksei Halavanau[2], Andrei Benediktovitch[9], Nina Rohringer[9,10], Matthias F. Kling[11,12], Claudio Pellegrini[2], Hitoki Yoneda[4], Uwe Bergmann[1*]

1. *Department of Physics, University of Wisconsin-Madison, Madison, WI, 53706, USA*
2. *SLAC National Accelerator Laboratory, Menlo Park, CA 94025*
3. *Stanford Synchrotron Radiation Lightsource, SLAC National Accelerator Laboratory, Menlo Park, CA, 94025, USA*
4. *Institute for Laser Science, The University of Electro-Communications, Chofu, Tokyo 182-8585, Japan*
5. *RIKEN SPring-8 Center, Sayo-cho, Sayo-gun, Hyogo 679-5148, Japan*
6. *Research Center for Precision Engineering, Graduate School of Engineering, Osaka University, Suita, Osaka 565-0871, Japan*
7. *Department of Advanced Materials Science, The University of Tokyo, 5-1-5 Kashiwanoha, Kashiwa, Chiba 277-0882, Japan*
8. *Japan Synchrotron Radiation Research Institute, Sayo-cho, Sayo-gun, Hyogo 679-5198, Japan*
9. *Center for Free-Electron Laser Science CFEL, Deutsches Elektronen-Synchrotron DESY, Notkestr. 85, 22607 Hamburg, Germany*
10. *Department of Physics, Universität Hamburg, Hamburg 22761, Germany*
11. *Stanford PULSE Institute. SLAC National Accelerator Laboratory, Menlo Park, California 94025, US*
12. *Department of Applied Physics, Stanford University. Stanford, California 94305, USA*

**Corresponding Authors**

*ubergmann@wisc.edu

*zabhari@wisc.edu

## Abstract:

X-ray free electron laser (XFEL) experiments using self-amplified spontaneous emission (SASE) pulses typically achieve temporal resolutions of order several femtoseconds, as the pulse duration puts a practical limit to pump-probe or probe-probe schemes. Even with the emerging capabilities to generate pulses with attosecond durations with new single-spike SASE schemes, direct access to attosecond electron dynamics remains an experimental challenge. Here we show how X-ray coherent attosecond pulse-pair spectroscopy (X-CAPPS) provides a powerful new approach to access the ultrashort time-delay window. Coherent attosecond pulse pairs with time delays varying from ~500 as to ~5 fs are generated with Cu $K\alpha_1$ stimulated X-ray emission from a gain medium pumped by intense SASE XFEL pulses. These pulse pairs are analyzed with two subsequent Bragg crystal spectrometers, and the resulting interference spectrum is captured on two sequential 2D image detectors encoding their time separation, relative amplitudes, and phases with high precision. X-CAPPS requires no split-and-delay X-ray optics, nor XFEL pulse modifications, making it broadly implementable across existing facilities. This technique enables the investigation of attosecond processes with Ångström resolution, providing a new tool for probing ultrafast dynamics across a wide range of atomic, molecular, and solid systems.

## Introduction:

Interference has been fundamental to scientific breakthroughs, from the Michelson-Morley experiment that laid the foundation for Einstein's special relativity to LIGO's gravitational wave detection[1–3]. This powerful property of the wave character of electromagnetic radiation, enables high precision measurements, facilitating the development of a vast number of techniques at many different wavelengths that continue to push research boundaries. Since Röntgen's discovery in 1895[4], X-rays have revolutionized science through their unique capabilities:

transmitting high energy radiation through materials for medical imaging; providing electronic structure information with elemental sensitivity through spectroscopy; and enabling atomic-scale structural determination through crystallography which enabled the elucidation of the DNA double-helix[5], arguably, one of the greatest discoveries of the 20th century.

Utilizing the interference at X-ray wavelengths has been a bedrock for precision measurements, most notably through Bragg and Laue diffraction. For interferometry, the short wavelengths and near-unity refractive indices pose technical challenges. In 1965, Bonse and Hart overcame these challenges when they constructed the first successful X-ray interferometer using a single silicon crystal in a triple-Laue geometry[6]. By maintaining perfect lattice coherence across all optical elements and exploiting transmissive geometry, they achieved Ångström-scale precision that has since enabled applications from phase contrast micrography to fundamental metrology[7,8]. Recent developments have sought to extend X-ray interferometry into new regimes, such as using Fano-resonance interference in nuclear forward scattering to achieve tunable phase sensitivity[9].

The advent of XFELs offers a new tool to transform the capabilities of interferometry, providing ultrafast probes with nanometer to Ångström spatial resolution and elemental specificity[10–15]. The intense, femtosecond X-ray pulses (~1-100 fs) have opened new experimental paradigms, and the desire to reach attosecond timescales where electronic dynamics unfold naturally, has driven growing theoretical and experimental efforts[16–21]. For instance, recent studies have demonstrated placing slotted spoiler foils upstream in XFEL accelerators can create temporally separated phase-locked X-ray pulse pairs for attosecond-scale interferometry[22]. Additionally, recent work on development of soft X-ray pump X-ray probe spectroscopy with sub-femtosecond temporal resolution[23], and its utilization in understanding the role of ultrafast nuclear dynamics versus inherent water structure in $1b_1$ peak splitting in water's oxygen K emission[24], have demonstrated the transformative potential of X-ray pulse pair spectroscopy in this temporal regime.

This pursuit of attosecond X-ray science has catalyzed the emergence of nonlinear X-ray physics, where techniques pioneered in the XUV domain are being extended to hard X-rays[25,26,27]. One of the largest hurdles impeding successful realization of nonlinear techniques with hard X-rays is the challenge of generating the necessary intense, coherent, ultrashort pulses with fixed relative phases. Promising efforts towards X-ray split-and-delay instrumentation are underway, but these systems face technical challenges to keep the relative phases of pulse pairs due to the mechanical instability and the requirement of substantial instrumentation footprints[28–30]. An equally promising and complementary alternative has emerged through XFEL-driven stimulated X-ray emission, which generates temporally coherent ultrafast X-ray pulses[31–36]. Our recent experiments demonstrated that $K\alpha_1$ emission from Manganese (Mn) gain media pumped by self-amplified spontaneous emission (SASE) XFEL pulses can generate coherent X-ray pulse pairs with 1.7 to 4.6 fs separations, evidenced by interference fringes in superfluorescence spectra[37], and generate pulses with durations on the order of 100 attoseconds[31]. Here we show that this approach of generating X-ray pulse pairs by stimulated X-ray emission can be employed to develop a new hard X-ray spectroscopy technique, X-ray coherent attosecond pulse-pair spectroscopy (X-CAPPS). Our X-CAPPS instrument is based on a copper (Cu) gain medium pumped with an intense and tightly focused SASE XFEL pulse to generate $K\alpha_1$ stimulated X-ray emission at 8.048 keV and two sequential Bragg spectrometers to analyze their resulting interference profiles. We observed 2D spectral-spatial profiles corresponding to X-ray pulse pairs with temporal spacing from 480 attoseconds to 5.2 fs, joining other efforts that have extended time resolved XFEL-based experiments to previously inaccessible time regimes[38–40]. The instrument is sensitive to small changes in relative amplitudes, phases, and temporal separations of the pulse pairs. Notably, X-CAPPS requires no modifications to existing XFEL infrastructure, enabling broad implementation across current facilities while providing a novel hard X-ray spectroscopy probe in the ultrafast time window currently unreachable by SASE XFEL pulses alone. Following is a discussion of the instrument and its capabilities, as well as the potential applications of X-CAPPS.

## Results:

### Principle of X-ray coherent attosecond pulse pair generation and detection

**Figure 1** illustrates the generation and detection scheme of coherent X-ray pulse pairs underlying X-CAPPS. The process starts when an intense SASE pump pulse generates a population inversion of 1s core hole excited states (**Fig. 1a)** in the Cu foil. If the pump pulse contains two strong temporal spikes, each spike can create an independent population inversion that generates separate superfluorescence bursts[37]. These bursts occur when Kα photons spontaneously emitted along the direction of beam propagation stimulate further emission deeper in the gain medium in the same direction, leading to a cascade of amplified spontaneous emission in the medium (**Fig. 1b**)[32,34]. At sufficient pump pulse intensities, or through the use of an externally provided second color seed pulse[33,35], the signal becomes very strong, leading to bursts of very short coherent X-ray pulses that have been described as superfluorescence[34]. The two coherent superfluorescence pulses exit the gain medium, as show in **Fig. 1d**, with time delays conditioned by their respective SASE spike separations, maintaining phase stability and remaining temporally separated due to their short durations. It is also possible to have multiple interference beyond just that of two pulses; see SI for more information.

The temporal relationship between these pulse pairs is extracted by observing their interference using a crystal spectrometer in Bragg geometry (**Fig. 1c,d**). Upon impinging on the Si analyzer crystal, the superfluorescence pulses are spectrally dispersed and each Fourier component is temporally stretched according to the spectral resolution of the analyzer to the point that they overlap temporally and interfere. Thus, resulting in a frequency-domain interference profile that reflects the temporal spacing of the superfluorescence pulses. On the spectral axis of the 2D profile, the fringe spacings, $\Delta E$, are inversely proportional to their time delays, $\Delta t$, given by $\Delta E = h/\Delta t$, where $h$ = 4.1356 fs·eV is the Planck constant. This defines the time delay between the two incoming superfluorescence pulses.

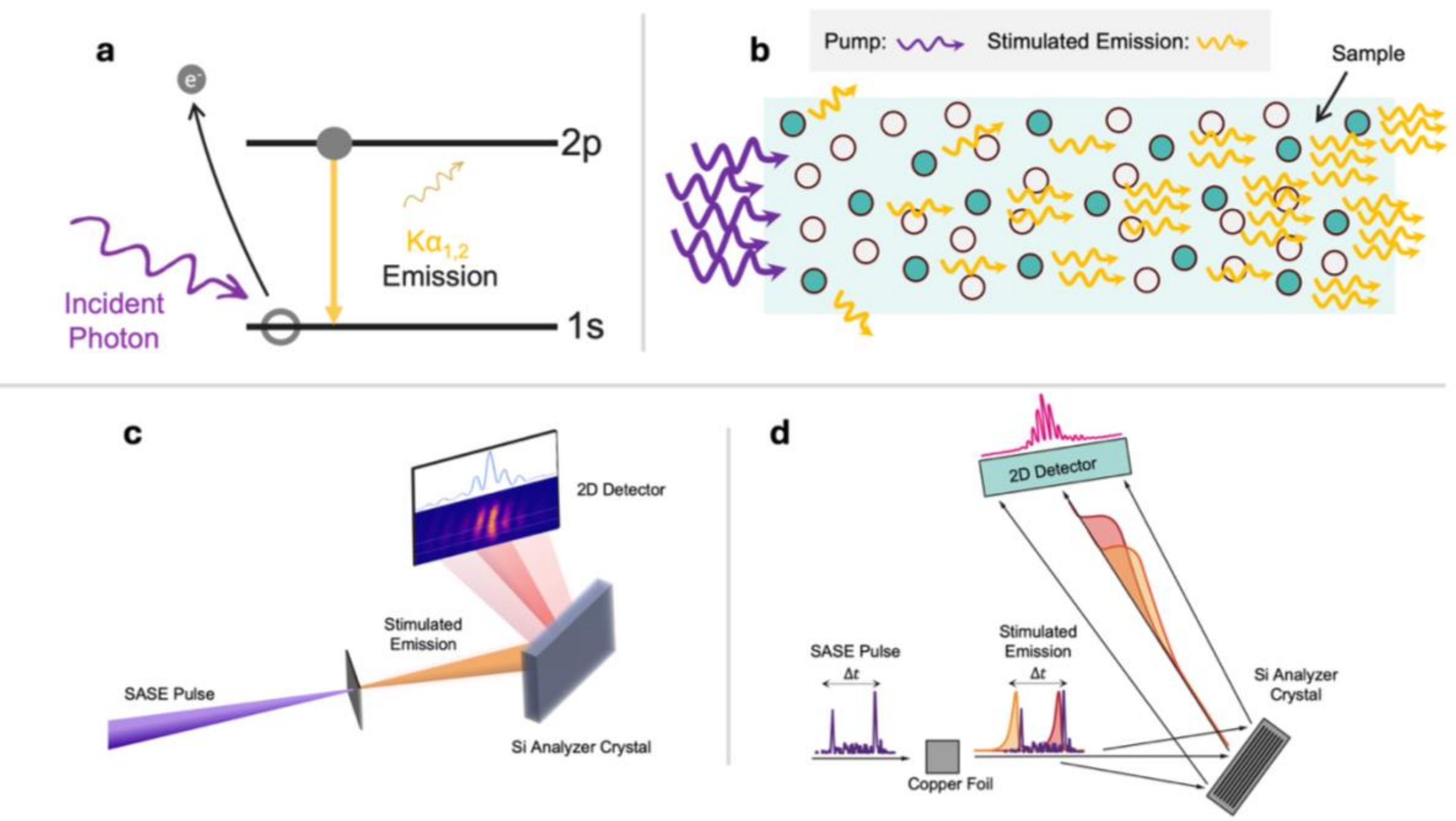

*Figure 1:* The concept of superfluorescence and its interference**.** (**a**) Level diagram showing $K\alpha_{1,2}$ emission resulting from the creation of a 1s core-hole. (**b**) Superfluorescence mechanism: The pump pulse (purple) creates 1s core

holes in the gain medium (Cu foil), and spontaneously emitted Kα photons (yellow) from the front of the medium stimulate further emission deeper within, creating a cascade. (**c**) Bragg spectrometer setup. (**d**) Interference process: A SASE pulse with two temporal spikes creates two superfluorescence events. The Si analyzer, set to the Cu Kα Bragg angle, rejects SASE photons while stretching the Fourier components of the superfluorescence pulses such that they overlap and interfere. The interference fringes are recorded with a 2D detector downstream of the analyzer with on spatial axis (horizontal) and one spectral axis (vertical) along the dispersive direction of the Bragg analyzer.

## Experimental parameters and setup

X-CAPPS is designed to capture the 2D profiles corresponding to the interference of coherent X-ray attosecond pulse pairs both before and after interaction with a sample, when one is present. In the absence of a sample, or when a sample affects both pulses identically, the recorded spectra are expected to be identical after correcting for transmission losses from the propagation through the first spectrometer and beamline components located between the two spectrometers. Any relative changes between the two pulses due to interaction with a sample will manifest as modifications in the spectrum resolved by the downstream spectrometer compared to that of the upstream spectrometer. For instance, if a sample's absorbance is altered between the first and the second pulse, the fringe contrast measured by the second analyzer will be reduced. Alternatively, changes in the relative phases of the two pulses caused by transmission through the sample will lead to a shift in the interference spectrum in energy. A related dual-spectrometer architecture has recently been employed for single-shot X-ray absorption spectroscopy (XAS) at XFELs, where upstream and downstream dispersive spectrometers are used to normalize stochastic SASE pulse fluctuations[41]. In that case, the comparison enables intensity-based absorption measurements, whereas in X-CAPPS the same general geometry is leveraged to resolve interference-encoded phase and amplitude changes with sub-femtosecond sensitivity.

To demonstrate X-CAPPS and its sensitivity, we performed experiments at the nanofocus instrument[42] in Experimental Hutch 5 (EH5) at beamline 3 at the SPring-8 Angstrom Compact free electron Laser (SACLA) in Hyogo, Japan[43,44]. We utilized 6 fs SASE pulses with ~200 μJ pulse energy that were tightly focused to a spot size of 100-150 nm in diameter using the beam line reflective focusing optics in Kirkpartrik-Baez geometry. The pulse intensity at the Cu gain medium was ~$10^{19}$-$10^{20}$ W/cm$^2$ with a central photon energy of 8.98keV and a bandwidth of 30 eV, tuned to be above the K edge of Cu. The KB optics resulted in an incoming beam with a 2.5 mrad vertical beam divergence and a 4 mrad horizontal beam divergence. It is important to note that the divergence of the superfluorescence signal is larger as it depends on the length of the gain medium with respect to the focus size[45].

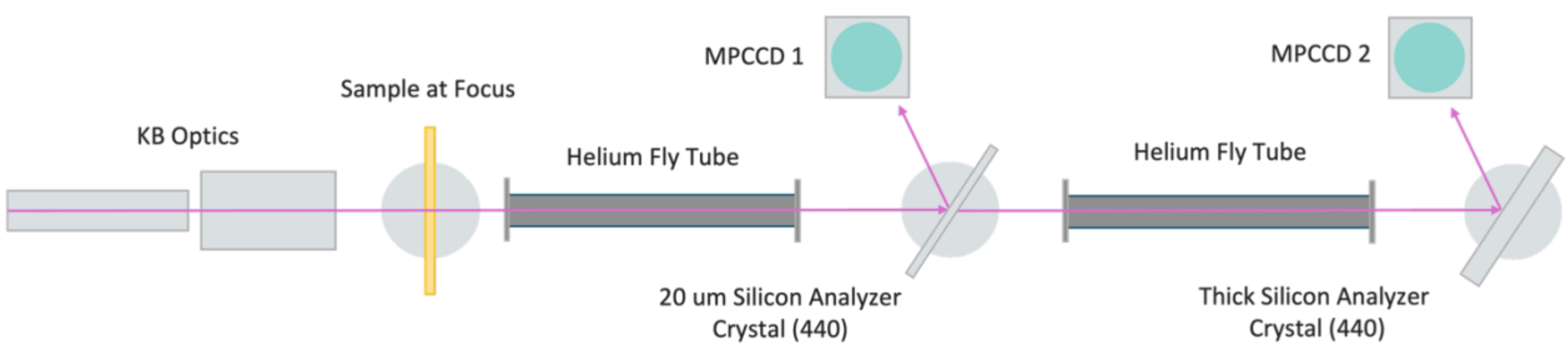

*Figure 2* The KB optics focus the XFEL beam onto the copper gain medium, generating superfluorescence. The interference of the superfluorescence is first measured by an upstream spectrometer after reflection from a 20 μm Si analyzer crystal. This thin crystal allows partial transmission, enabling the interference pattern to be recorded again on a downstream spectrometer after a second reflection from a flat Si analyzer crystal.

The dual-spectrometer setup is detailed in **Fig. 2**, which illustrates the main components of the X-CAPPS instrument. The superfluorescence pulse pairs were analyzed using two sequentially aligned Bragg spectrometers. Each spectrometer consists of a Si analyzer crystal and a 2D multi-port charge-coupled device (MPCCD)[46] with a 50 μm pixel size made up of 1024 pixels vertically and 512 pixels horizontally. The flat analyzer crystals were utilized in a vertical Bragg scattering geometry, where the slightly divergent superfluorescence signal is dispersed onto the 2D detector with one vertical ‘spectral’ axis and one horizontal ‘spatial/angular’ axis. The upstream spectrometer employs a thin, largely transparent, 20 μm Si crystal analyzer (Norcada Inc.), and the downstream spectrometer, a thick Si(440) crystal analyzer using the same crystal orientation. The analyzers are set at a Bragg angle $\theta_B$ = 53.353° for the central pulse direction, corresponding to the Cu $K\alpha_1$ line at E = 8.048 keV, and the center of the MPCCDs are positioned at $2\theta_B$. The 20 μm thin analyzer crystal of the upstream spectrometer provides approximately 70% reflectivity due to the short extinction length ($\xi_\sigma = 7.153$ μm, $\xi_\pi = 24.879$ μm) at the Si(440) Bragg reflection, while allowing for approximately 90% transmission in the off-Bragg condition. Given the large vertical and horizontal divergence of the superfluorescence signal (~8 mrad) compared to the Darwin width of the Si(440) reflection at 8.048 keV of ~9 μrad ($\omega_\sigma$ =14.31 μrad, $\omega_\pi$ = 4.11 μrad), there is an advantage of introducing a slight mismatch between the Bragg angles of the upstream and downstream spectrometers. An example is shown in the DuMond diagram in **Fig. 3b**, where a mismatch of 14 μrad between the two analyzers is depicted, ensuring that the first analyzer does not cut out the exact phase space that would block the second analyzer to receive a signal. It is important to note that this DuMond diagram represents only one horizontal slice of the phase space. More information on the DuMond diagram can be found in the SI. We used an optical pre-alignment which did not have the precision of encountering the undesirable situation where the perfect Bragg angle and horizontal alignment of the two analyzers would lead to an exact overlap of the Bragg-reflected phase space and consequent loss of signal in the second analyzer. We also did not observe any effects of this slight misalignment on the measured 2D profiles, confirming that the spectral changes of each superfluorescence pulse does not have a strong correlation with the emission direction.

The spectral resolution is given by the convolution of analyzer resolution given by the Si (440) Darwin width (~9 μrad = 120 meV) and geometric resolution set by the detector pixel size $p$ and its distance $L$ from the gain medium, according to $\Delta E = \frac{p}{L}\frac{E}{tan\theta_B}$. For our setup the overall spectral resolution is ~0.3 eV for the upstream spectrometer and ~0.15eV for the downstream spectrometer. The upstream spectrometer is 1.03 m from the source of the pulse pair, and the downstream spectrometer is 2.1 m from the source of the pulse pair.

The accessible energy range $\Delta E$ of our spectrometer is defined by the photon energy $E$ and angular divergence $\Delta\theta$ of the superfluorescence signal, and the Bragg angle $\theta_B$, through the equation $\Delta E = E \cdot \cot(\theta_B) \cdot \Delta\theta$, which is the derivative of the Bragg's law equation. The divergence of the superfluorescence signal is approximately determined by the ratio of beam size to gain length, with a lower limit set by the gain medium thickness[45]. For our 25 μm Cu foil gain medium, we estimate the superfluorescence divergence to be at least $\Delta\theta \approx$ 200 nm/25 μm = 8 mrad. This translates to a spectral range of $\Delta E$ = 47 eV for the upstream spectrometer, which defines the accessible spectral range of the entire X-CAPPs instrument. This setup and method can be used with single spike split and delay techniques already available[38,47,48].

The temporal separation, $\Delta t$, of the pulse pairs that can be resolved with X-CAPPS is directly linked to the temporal characteristics of the incoming SASE pulse. The upper limit is conditioned by the overall temporal length of the incoming SASE pulse. With the 6fs pulses provided at SACLA, the upper end of the accessible time window is just below 6fs, corresponding to a fringe energy spacing of 0.7eV. The finest fringe spacing we observed was 0.85eV in a previous experiment.  With longer SASE pulses, longer delays can be achieved. The theoretical lower limit is also conditioned by the characteristics of the SASE pulse, but in practice is limited by the divergence of the superfluorescence. Within one SASE pulse, two temporal spikes can occur as near to one another as 100 – 200 as, given by the coherence time of the FEL[49]. However, the divergence of the superfluorescence sets the accessible energy window for the Bragg spectrometer. For our experiment, based on the ~8 mrad divergence, we have ~47 eV energy window. To know that we have observed an interference, we need to resolve at least three regularly spaced peaks within that energy window, which sets our lower limit to around 275 as, which would correspond to a fringe energy spacing of 15eV. The largest fringe spacing we observed was 8.7eV, also observed in a previous experiment.

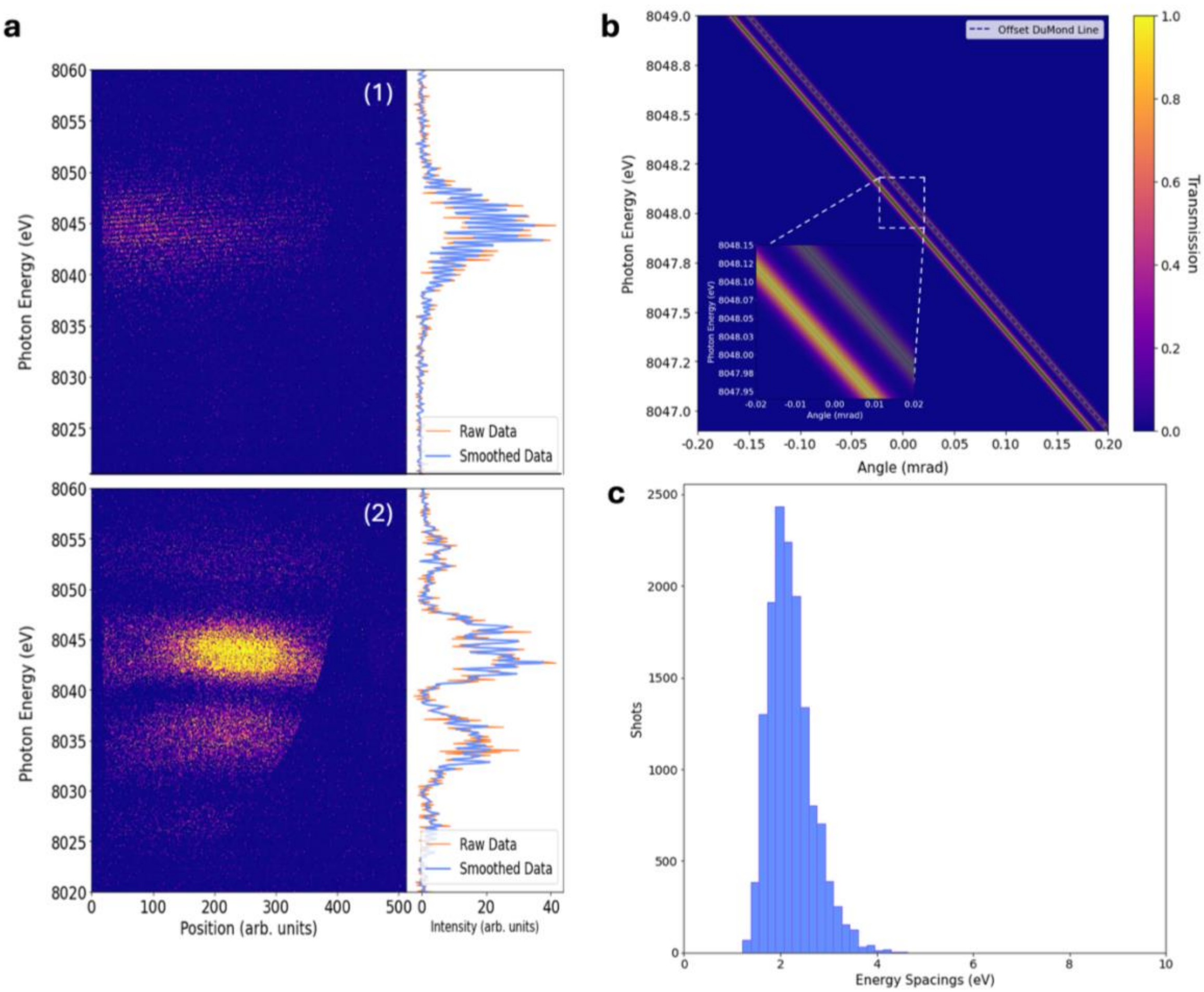

*Figure 3:* (**a**) The two 2D emission profiles illustrate the spectral and temporal range captured by the spectrometer. The smallest fringes (1) of 0.85eV spacing correspond to a temporal spacing of 4.89 femtoseconds, while the largest fringes (2) of 8.7eV spacing corresponds to 475 attoseconds. (**b**) The DuMond diagram illustrates the phase space of the central beam cut out by each spectrometer, assuming a 14 μrad offset between the Bragg angles of the upstream and downstream analyzers. (**c**) The histogram represents the variety and frequency of fringe spacings observed in the commissioning experiment.

Since the temporal spacing between individual spikes within a SASE pulse cannot be controlled, the inter-pulse delay (and therefore the fringe spacing) varies stochastically from shot to shot. Experiments exploiting this technique must therefore acquire large datasets to accumulate sufficient statistics across the full range of physically relevant delays. The statistics for this experiment are shown in **Fig. 3c**. The method of selecting fringe shots is outline in the SI.

## Instrument Precision

Ideally, the two spectrometers can record identical 2D profiles when no sample is inserted in the beam path and transmission losses, alignment, and geometry effects are accounted for. The precision due to these various effects was evaluated using three complementary methods: (1) visual inspection through contour overlay, (2) Pearson correlation of one-dimensional (1D) spectral projections, and (3) simulation evaluations of sample induced changes.

Beginning with a visual inspection, fringe contours were extracted from the upstream spectrometer image using SAOImageDS9[50] and overlaid onto the downstream image. As illustrated in **Fig. 4**, the contours remained well-aligned even for complex fringe structures, confirming the spatial consistency of the resolved emission.

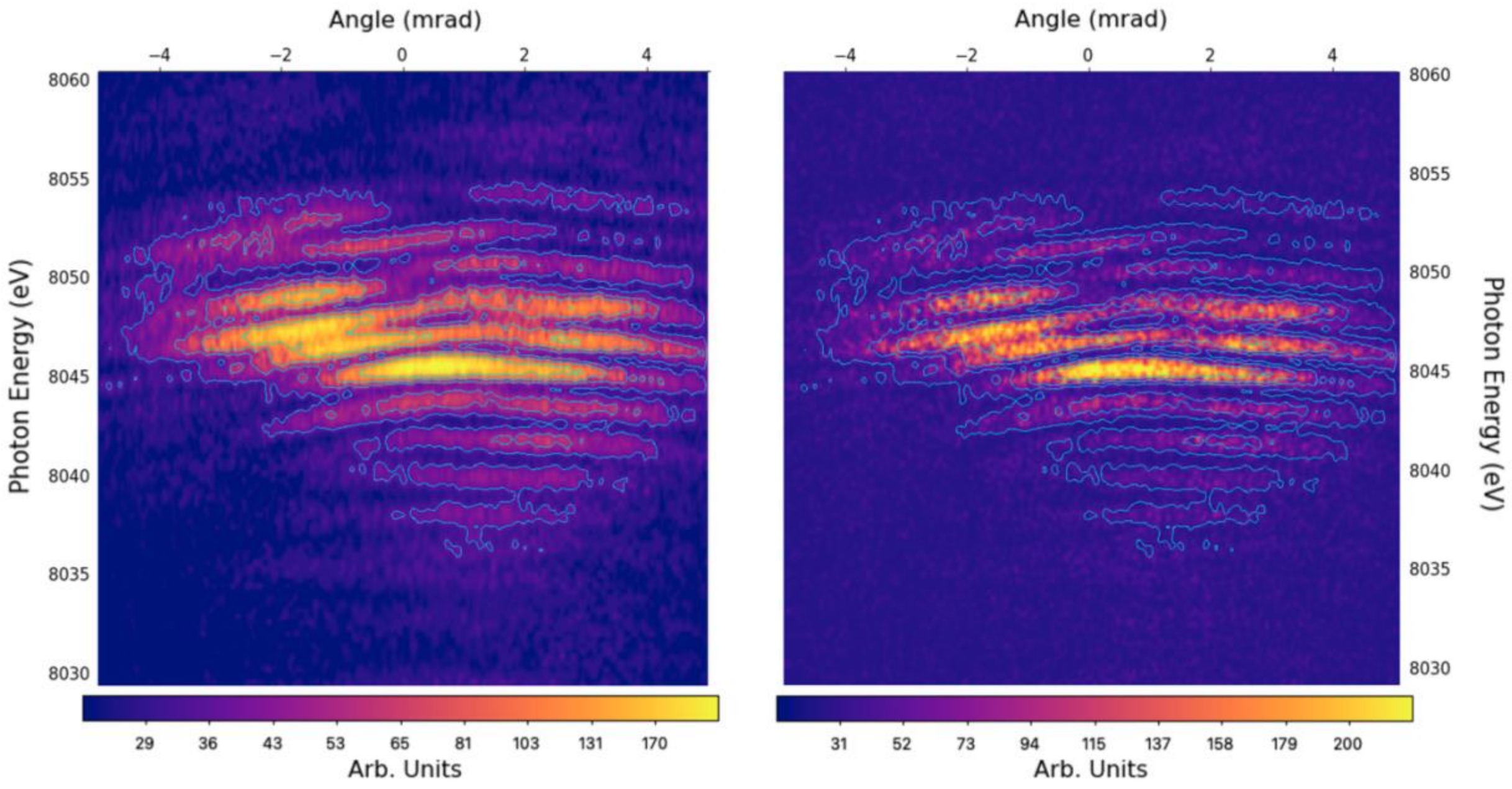

*Figure 4*: Contours were extracted from the upstream spectrometer image using SAOImageDS9 and overlaid onto the downstream spectrometer image to visually assess the ability to resolve identical images on both spectrometers. Selecting a valid fringe with a complex structure, it is evident that image reproducibility between the two spectrometers is high.

For applications involving spectroscopy and interferometry with the X-CAPPS instrument, rather than analyzing the 2D profiles, more information can be gathered from comparing 1D cuts of the dispersion to determine if there is any modulation in the downstream spectrum. To quantify spectral similarity, we computed the Pearson correlation coefficient between upstream and downstream 1D projections for each valid spectrum as seen in **Fig. 5b**. A spectrum was considered valid if it exceeded a defined intensity threshold and did not correspond to a misfire; both interference and non-interference spectra were included. Of the 30,000 valid shots analyzed, fewer than 0.1% exhibited a correlation coefficient of $R = 0.96$, while 98% showed values of $R > 0.99$, indicating a high degree of reproducibility across the two spectrometers.

Beyond reproducibility, we also evaluated the instrument's precision through simulation. As discussed earlier, the most prominent observables are expected to be either contrast changes or phase shifts. Sensitivity to absorption-based modulations is relatively limited; only large reductions in pulse intensity (greater than ~20%) can be resolved given the pixel size and spectrometer resolution.

In contrast, phase shifts can be detected with far greater precision. When the superfluorescence spectrum is defined by a series of narrow, regularly spaced spectral features, the measurement becomes analogous to a frequency comb, in which even small relative shifts produce a pronounced change in the resulting comparison of the interference spectra. Simulations of typical fringe spectra, modeled with 2 eV spacings, shown in **Fig. 5a**, demonstrate that relative phase shifts on the order of ~2% can be resolved. While the X-CAPPS instrumentation is not yet optimized, our results highlight the technique's capacity for high-precision measurements and its significant potential for further improvement.

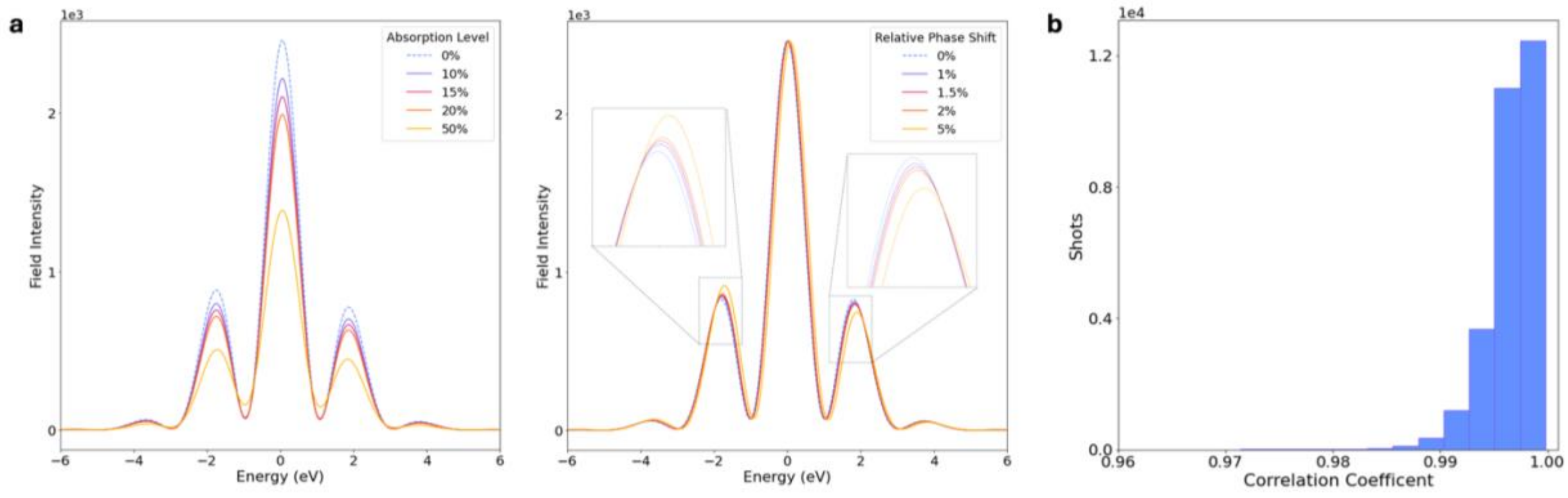

*Figure 5*: (**a**) Sensitivity of the X-CAPPS instrument to sample-induced absorption and phase shifts. The left panel shows simulated downstream spectra where the intensity of one of the two pulses is reduced, demonstrating what would happen if there was sample induced absorption. The right panel, similarly shows how a relative phase shift between the two pulses would like when compared to the reference spectrum. (**b**) One-dimensional Pearson correlation analysis of upstream and downstream spectra. Correlations between one-dimensional projections of the recorded two-dimensional spectra are shown for a random selection of 1,400 shots out of ~14,000 shots.

**Discussion and Outlook:**

X-CAPPS combines nonlinear X-ray pulse generation with interference detection into a deployable framework for ultrafast time-dependent X-ray measurements, providing a new approach to probing the electronic and structural dynamics on the fastest time scales. As XFEL facilities continue to advance toward shorter pulses and improved coherence, the combination of intrinsic phase sensitivity, spectral interferometry, and compatibility with existing infrastructure positions X-CAPPS as a promising platform for future studies of ultrafast phenomena, including dense plasma dynamics, coherent electronic motion, and nonlinear X-ray interactions.

A possible application of X-CAPPS is the investigation of dense laser-generated plasmas through measurements of plasma-induced spectral phase shifts and fringe visibility changes. By comparing the spectral interference profiles recorded before and after interaction with the plasma, changes in fringe position, spacing, and contrast could provide simultaneous sensitivity to dispersive phase shifts, absorption, and coherence loss within the plasma. Unlike conventional X-ray phase-contrast imaging techniques, which primarily probe spatial gradients in electron density, the interferometric method is directly sensitive to the complex refractive response encoded in the X-ray field itself. Furthermore, because the spectral fringe spacing encodes the temporal separation of the pulse pair on a shot-to-shot basis, statistical analysis of many single-shot interference profiles may enable access to ultrafast plasma evolution without requiring precise external timing control of the XFEL pulse structure.

The performance reported here represent the operating conditions of the first experiments that can be further optimized in a straightforward way. Improvements in analyzer alignment, spectral resolution, gain-medium optimization, and pulse-pair generation statistics are expected to enhance the sensitivity of X-CAPPS by an order of magnitude. Extensions to other gain medium materials, including heavier elements with L emission lines that exhibit shorter core-hole lifetimes, will allow access to different photon energies and even shorter time delays. X-CAPPS can also be integrated into existing XFEL split-and-delay instruments to enable hybrid schemes combining coarser multi femtosecond temporal control with attosecond interference sensitivity. Such developments will position X-CAPPS as a powerful and versatile complement to other emerging attosecond XFEL capabilities.

**Acknowledgements:**

We acknowledge the support from the SACLA accelerator groups and their technical and engineering staff. This work is supported by the US Department of Energy (DOE), Office of Science, Basic Energy Sciences (BES) DE-SC-0023585 (A.H., U.B. and C.P.); Office of Basic Energy Sciences (OBES), Division of Chemical Sciences, Geosciences and Biosciences (CSGB) under contract nos. DE-SC0023270 (T.M.L., Z.A. and U.B.). The experiment at SACLA was performed with the approval of the Japan Synchrotron Radiation Research Institute (proposal nos. 2022B8002, 2023B8053, 2023B8039, 2024A8055). We acknowledge the SACLA Graduate Student Support Program (Z.A). We acknowledge JPSJ KAKENHI for grant nos. 19K20604, 22KK0233, 23K25131, 24K21199 (II) and 22K1813 (T.O.). I.I. acknowledges the financial support from JST PRESTO (JPMJPR24J1). SSRL Structural Molecular Biology Program is supported by the DOE Office of Biological and Environmental Research and the National Institutes of Health, National Institute of General Medical Sciences (including P41GM103393) (T.K.). The contents of this publication are solely the responsibility of the authors and do not necessarily represent the official views of NIGMS or NIH (T.K.).

**References:**

1. *Basics of Interferometry*. (Elsevier, 2007). doi:10.1016/B978-0-12-373589-8.X5000-7.
2. Michelson, A. A. & Morley, E. W. On the relative motion of the Earth and the luminiferous ether. *Am. J. Sci.* **s3-34**, 333–345 (1887).
3. Einstein, A. *The Collected Papers of Albert Einstein*. (Princeton university press, Princeton, 1987).
4. On a New Kind of Rays. *Nature* **53**, 274–276 (1896).
5. Franklin, R. E. & Gosling, R. G. Molecular Configuration in Sodium Thymonucleate. *Nature* **171**, 740–741 (1953).
6. Bonse, U. & Hart, M. An X-ray interferometer. *Appl. Phys. Lett.* **6**, 155–156 (1965).
7. Hart, M. An $aring$ngstr$ouml$m ruler. *J. Phys. Appl. Phys.* **1**, 1405–1408 (1968).
8. Massa, E., Sasso, C. P. & Mana, G. The Measurement of the Silicon Lattice Parameter and the Count of Atoms to Realise the Kilogram. *MAPAN* **35**, 511–519 (2020).
9. Heeg, K. P. *et al.* Interferometric phase detection at x-ray energies via Fano resonance control. *Phys. Rev. Lett.* **114**, 207401 (2015).
10. Bostedt, C. *et al.* Linac Coherent Light Source: The first five years. *Rev. Mod. Phys.* **88**, 015007 (2016).
11. Bucksbaum, P. H. & Berrah, N. Brighter and faster: The promise and challenge of the x-ray free-electron laser. *Phys. Today* **68**, 26–32 (2015).
12. Zhu, D. & Reis, D. A. Attosecond X-ray laser vision: High-energy photon sources. *Nat. Photonics* **18**, 1232–1233 (2024).

13. Bergmann, U. *et al.* Using X-ray free-electron lasers for spectroscopy of molecular catalysts and metalloenzymes. *Nat. Rev. Phys.* **3**, 264–282 (2021).

14. Ishikawa, T. Early Days of SACLA XFEL. *Photonics* **9**, (2022).

15. Emma, P. *et al.* First lasing and operation of an ångstrom-wavelength free-electron laser. *Nat. Photonics* **4**, 641–647 (2010).

16. Driver, T. *et al.* Attosecond delays in X-ray molecular ionization. *Nature* **632**, 762–767 (2024).

17. Robles, R. *et al.* Broadband hard X-ray attosecond pulses from extremely chirped electron beams. Preprint at https://doi.org/10.48550/ARXIV.2604.09969 (2026).

18. Berrah, N. *et al.* Attosecond X-ray sources, methods, and applications at present and future free-electron lasers: tutorial. *Adv. Opt. Photonics* **17**, 623 (2025).

19. Inoue, I. *et al.* Experimental demonstration of attosecond hard X-ray pulses. Preprint at https://doi.org/10.48550/ARXIV.2506.07968 (2025).

20. Yan, J. *et al.* Terawatt-attosecond hard X-ray free-electron laser at high repetition rate. *Nat. Photonics* **18**, 1293–1298 (2024).

21. Duris, J. *et al.* Tunable isolated attosecond X-ray pulses with gigawatt peak power from a free-electron laser. *Nat. Photonics* **14**, 30–36 (2020).

22. Reiche, S. *et al.* A perfect X-ray beam splitter and its applications to time-domain interferometry and quantum optics exploiting free-electron lasers. *Proc. Natl. Acad. Sci.* **119**, e2117906119 (2022).

23. Guo, Z. *et al.* Experimental demonstration of attosecond pump–probe spectroscopy with an X-ray free-electron laser. *Nat. Photonics* **18**, 691–697 (2024).

24. Li, S. *et al.* Attosecond-pump attosecond-probe x-ray spectroscopy of liquid water. *Science* **383**, 1118–1122 (2024).

25. Young, L. *et al.* Roadmap of ultrafast x-ray atomic and molecular physics. *J. Phys. B At. Mol. Opt. Phys.* **51**, 032003 (2018).

26. Kowalewski, M., Fingerhut, B. P., Dorfman, K. E., Bennett, K. & Mukamel, S. Simulating Coherent Multidimensional Spectroscopy of Nonadiabatic Molecular Processes: From the Infrared to the X-ray Regime. *Chem. Rev.* **117**, 12165–12226 (2017).

27. Mukamel, S., Healion, D., Zhang, Y. & Biggs, J. D. Multidimensional Attosecond Resonant X-Ray Spectroscopy of Molecules: Lessons from the Optical Regime. *Annu. Rev. Phys. Chem.* **64**, 101–127 (2013).

28. *Synchrotron Light Sources and Free-Electron Lasers: Accelerator Physics, Instrumentation and Science Applications*. (Springer International Publishing, Cham, 2020). doi:10.1007/978-3-030-23201-6.

29. Mandal, A., Sidhu, M. S., Rost, J. M., Pfeifer, T. & Singh, K. P. Attosecond delay lines: design, characterization and applications. *Eur. Phys. J. Spec. Top.* **230**, 4195–4213 (2021).

30. Zhu, D. *et al.* Development of a hard x-ray split-delay system at the Linac Coherent Light Source. in (eds Tschentscher, T. & Patthey, L.) 102370R (Prague, Czech Republic, 2017). doi:10.1117/12.2265171.

31. Linker, T. M. *et al.* Attosecond inner-shell lasing at ångström wavelengths. *Nature* **642**, 934–940 (2025).

32. Kroll, T. *et al.* Stimulated X-Ray Emission Spectroscopy in Transition Metal Complexes. *Phys. Rev. Lett.* **120**, 133203 (2018).

33. Doyle, M. D. *et al.* Seeded stimulated X-ray emission at 5.9 keV. *Optica* **10**, 513 (2023).

34. Rohringer, N. *et al.* Atomic inner-shell X-ray laser at 1.46 nanometres pumped by an X-ray free-electron laser. *Nature* **481**, 488–491 (2012).

35. Yoneda, H. *et al.* Atomic inner-shell laser at 1.5-ångström wavelength pumped by an X-ray free-electron laser. *Nature* **524**, 446–449 (2015).

36. Kroll, T. *et al.* Multiplet lines in seeded stimulated Mn K α 1 x-ray emission. *Phys. Rev. Appl.* **25**, L051006 (2026).

37. Zhang, Y. *et al.* Generation of intense phase-stable femtosecond hard X-ray pulse pairs. *PNAS* **119**, (2022).

38. Sun, Y., Li, H., Ichii, Y. & Zhu, D. An ultrastable hard x-ray attosecond split-delay line. Preprint at https://doi.org/10.48550/arXiv.2505.06865 (2025).

39. Osaka, T. *et al.* Hard x-ray intensity autocorrelation using direct two-photon absorption. *Phys. Rev. Res.* **4**, L012035 (2022).

40. Li, H. *et al.* Generation of highly mutually coherent hard-x-ray pulse pairs with an amplitude-splitting delay line. *Phys. Rev. Res.* **3**, 043050 (2021).

41. Harmand, M. *et al.* Single-shot X-ray absorption spectroscopy at X-ray free electron lasers. *Sci. Rep.* **13**, 18203 (2023).

42. Yumoto, H. *et al.* Nanofocusing Optics for an X-Ray Free-Electron Laser Generating an Extreme Intensity of 100 EW/cm2 Using Total Reflection Mirrors. *Appl. Sci.* **10**, 2611 (2020).

43. Yabashi, M., Tanaka, H. & Ishikawa, T. Overview of the SACLA facility. *J. Synchrotron Radiat.* **22**, 477–484 (2015).

44. Ishikawa, T. *et al.* A compact X-ray free-electron laser emitting in the sub-ångström region. *Nat. Photonics* **6**, 540–544 (2012).

45. Kroll, T. *et al.* Observation of Seeded Mn Kβ Stimulated X-Ray Emission Using Two-Color X-Ray Free-Electron Laser Pulses. *Phys. Rev. Lett.* **125**, (2020).
46. Kameshima, T. *et al.* Development of an X-ray pixel detector with multi-port charge-coupled device for X-ray free-electron laser experiments. *Rev. Sci. Instrum.* **85**, 033110 (2014).
47. Roling, S. *et al.* A split- and delay-unit for the European XFEL. in (eds Tschentscher, T. & Tiedtke, K.) 87781G (Prague, Czech Republic, 2013). doi:10.1117/12.2027029.
48. Osaka, T. *et al.* Wavelength-tunable split-and-delay optical system for hard X-ray free-electron lasers. *Opt. Express* **24**, 9187 (2016).
49. Zhou, G. *et al.* Attosecond Coherence Time Characterization in Hard X-Ray Free-Electron Laser. *Sci. Rep.* **10**, 5961 (2020).
50. Smithsonian Astrophysical Observatory. SAOImage DS9: A utility for displaying astronomical images in the X11 window environment. *Astrophys. Source Code Libr.* ascl:0003.002 (2000).

## Supplemental Information

# Fringe Finding

### Algorithm

Based on the SciPy function, *scipy.signal.find_peaks*, the fringe finding algorithm is written such that it selects out single shots that resemble fringe spectra that would be useful for the interferometry application we describe in the main work. See the next section for how we define usable shots. Each shot is passed through a series of gates to determine if it is a valid and usable shot.

*Tilt*

The first gate determines whether the overall emission profile is tilted. For each shot, a 1D projection of the fringe region is computed by summing pixel intensities along lines at each angle, stepping from 0° to 180°. The maximum value of each 1D projection is recorded as a measure of how concentrated the intensity is along that projection direction. When the maximum projection strength occurs near 90°, the fringes are oriented horizontally — since projecting vertically (90°) across horizontal fringes produces the most peaked intensity distribution. This is essentially a Radon transform. Shots where the dominant projection angle deviates from 90° by more than 3° are rejected. In the figure shown below, the left shot would be considered tilted and rejected, whereas the right image would be accepted at this gate.

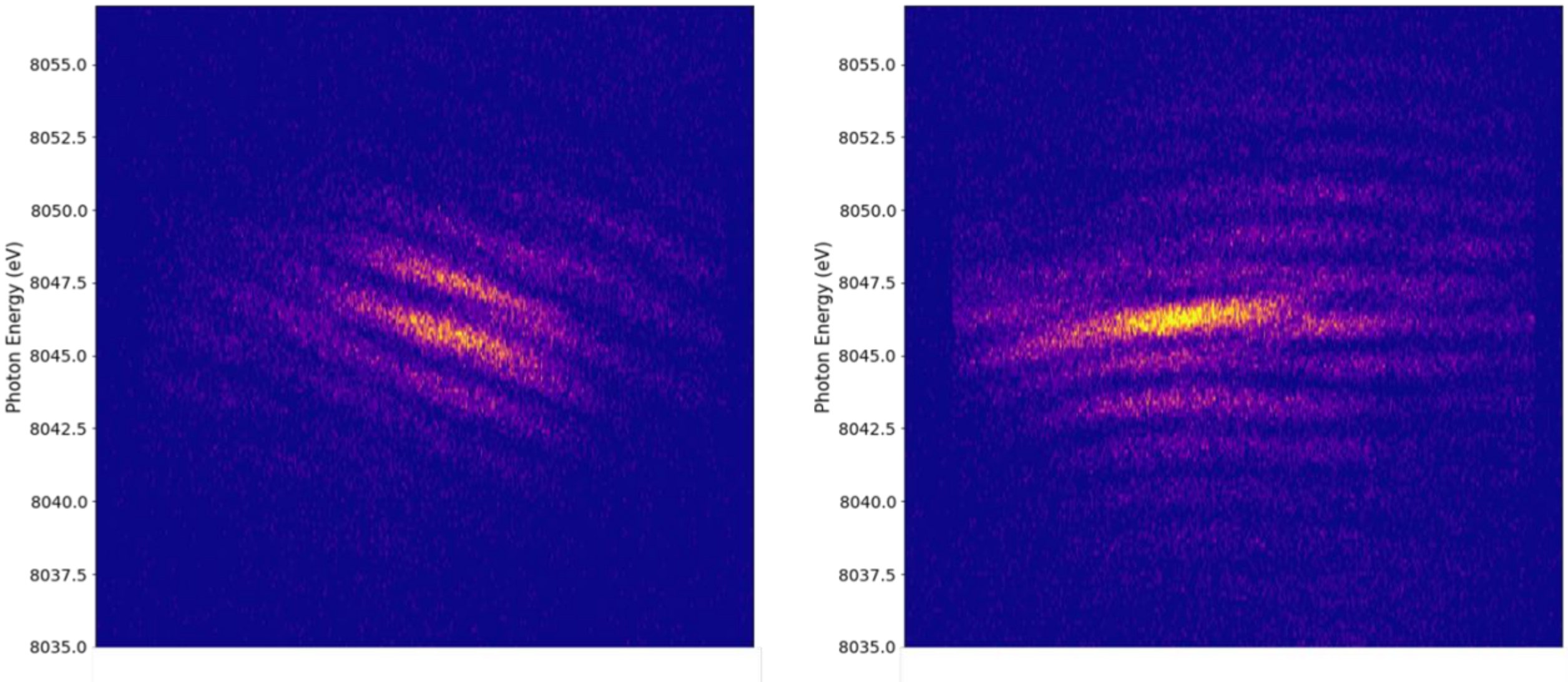

*S6:* Tilted vs. Untiled Fringes. The left image shows what we consider to be a tilted fringe. The right image is a shot that would pass at this gate, but not necessarily all others.

*Spatial Consistency*

This gate is to determine if there exists and spatial inhomogeneities in the interference profiles. Starting from the most intense point of the image, every ±5 pixels, a 5-pixel slice is taken and each of these slices is then smoothed to suppress high-frequency noise with a Savitzky–Golay filter, windowing every 7$^{th}$ point, interpolating to a 4$^{th}$ order polynomial. They are then passed through the peak finder and the determined energy spacings for each projection is compared to all the others. If the relative standard deviation falls below the threshold (0.12), determined through optimization of the algorithm, these shots are considered valid.

*Valley Depth*

This gate is to ensure the selection of high contrast interference profiles. First, a 5-pixel slice is taken at the most intense point within the 2D image. This projection is then passed through the peak finder post smoothing. For each consecutive pair of peaks detected, the valley is pinpointed by the minimum value that exists between the two peaks. The depth of the valley is then computed as follows:

$$Depth = \frac{Average\ intensity\ of\ first\ peak - Valley\ intensity}{Average\ intenisity\ of\ second\ peak - Valley\ intensity}$$

If the depth value is 0, that means the valley and peaks are the same height, and if the depth value is 1, that means that the valley itself is at zero indicating perfect contrast. All the depth values are averaged and compared to the threshold value (0.2), determined through optimization of the algorithm and accounting for the line spread function of the detector.

*Spacing Regularity*

This gate is to ensure the selected interference profiles exhibit regular spacing between all detected peaks. Based on the preprocessing done in the valley depth gate (smoothed, passed to peak finder), the average spacings between peaks are compared to ensure regularity based on a relative standard deviation threshold (0.5) determined by optimization. The imposes smaller acceptable deviations for finer interference patterns and larger deviations for larger patterns.

*Peak Requirement Gate*

This gate is to simply ensure there are multiple peaks, indicative of an interference event.

While this algorithm serves the purposes of the present study, the complexity of the selection criteria highlights a natural opportunity for image recognition machine learning, which will be explored in future work.

**Usable Shots**

We define usable shots as high-contrast, interference patterns oriented perpendicular to the energy axis. The majority of shots contain some form of spectral interference alongside spatial inhomogeneities that arise from additional nonlinear phenomena driven at the high intensities required to produce superfluorescence. The presence of multiple competing phenomena within a single shot would make it difficult to isolate the origin of any observed spectral features. We therefore select only shots in which the fringe pattern is consistent with the interference of two superfluorescence bursts, and exclude those exhibiting signatures of additional nonlinear processes. Below are examples of ‘usable’ and ‘unusable’ points. The top row represents unusable shots and the bottom row are usable shots.

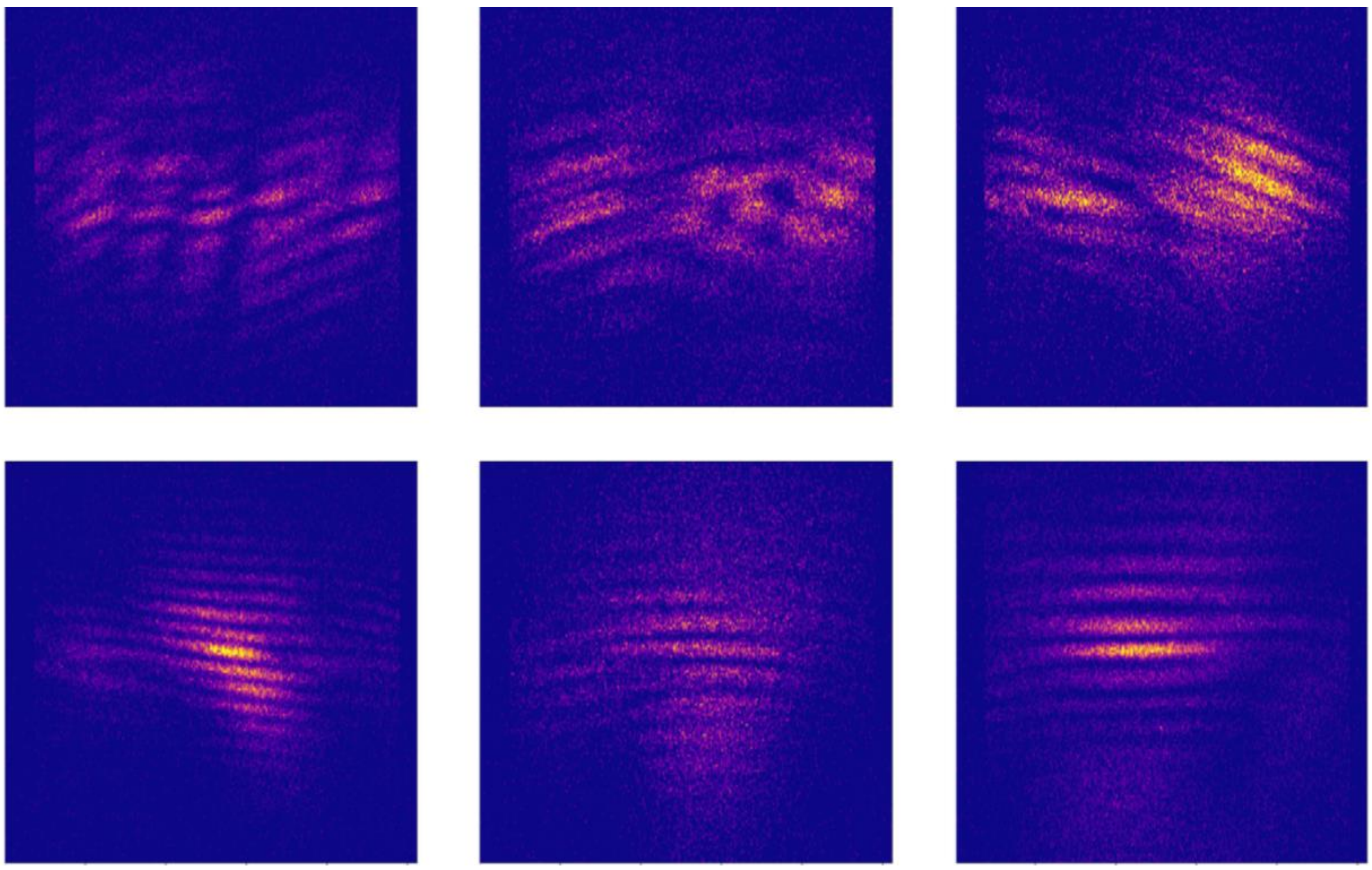

*S7:* Usable vs Unusable Shots. The top row contains a series of shots that would be classified as unusable. In addition to their spectral inhomogeneities, they contain multiple spatial inhomogeneities that would make it difficult to discern specific sample induced changes. The bottom row contains a series of shots that would make it through the fringe finding algorithm and would be deemed as usable for potential applications of the X-CAPPS instrument.

# Simulations

## DuMond Diagrams

To visualize the effect of small relative angular offsets between the two analyzer crystals, we constructed DuMond diagrams in joint phase space. The simulations were performed over ±2 mrad and 8035–8060 eV using a dense numerical grid. The central energy–angle dispersion for each analyzer was obtained from a linearized form of Bragg's law valid for small angular deviations about the nominal Bragg angle. In this approximation, small angular offsets produce a linear energy shift proportional to $E_B \cot \theta_B$, defining the DuMond line in phase space.
The two analyzers were modeled with slightly different Bragg parameters (8048.0 eV at −53.353° and 8048.1 eV at −53.35379°), corresponding to a small relative angular offset, ~14 urad. The intrinsic Darwin width of the reflection (9 μrad) was converted into an effective angular acceptance of 11 μrad to account for projection effects at the operating geometry. This effective angular width was mapped into an equivalent energy bandwidth via the linearized Bragg equation, and each analyzer's reflectivity was represented by a Gaussian transmission profile centered on its DuMond line. The overlaid two-dimensional transmission maps illustrate that, despite the small angular offset, the corresponding phase-space slices remain effectively non-overlapping within the relevant angular window. As a result, each analyzer samples a distinct region of angle–energy phase space, enabling the two spectrometers to produce indistinguishable two-dimensional detector images even in the presence of small relative angular misalignments.

## Additional Observations

### Indications of higher-order interference

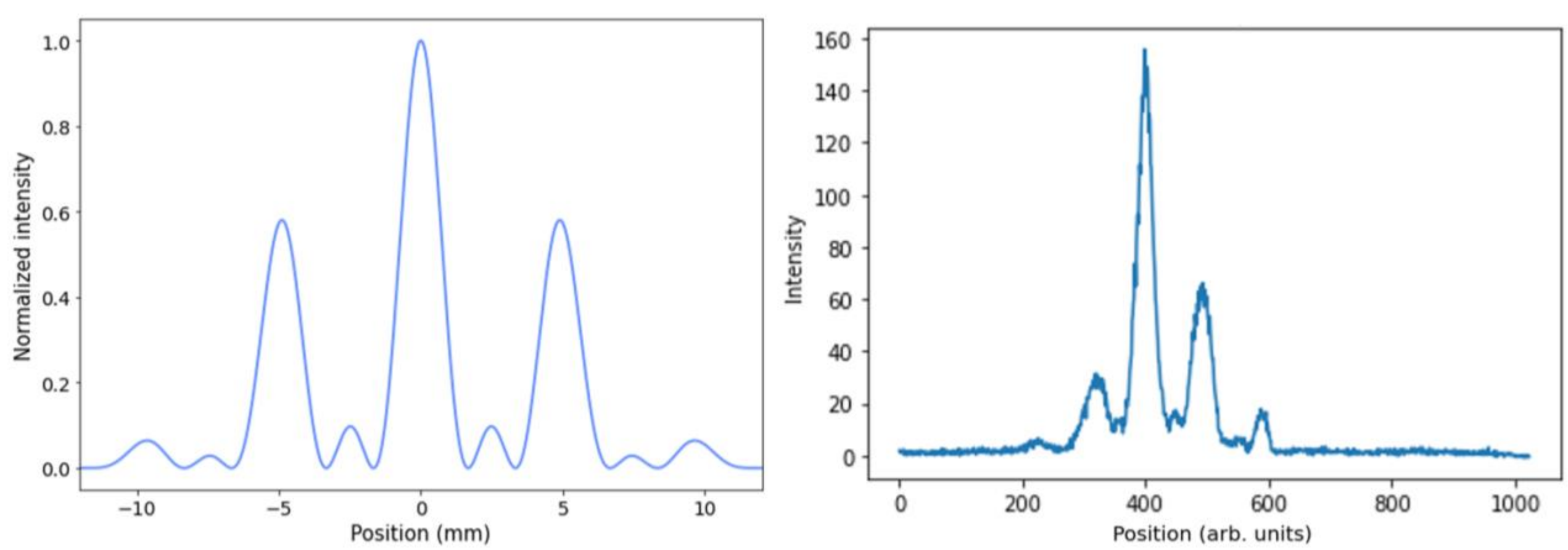

*S8:* Multiple Interference. The left image is a simulated spectrum corresponding to the result of 3-slit diffraction to model the right image of experimental data. The right image suggests it is possible for there to be instances of more than two pulse pairs being generated in a single shot.

In addition to the dominant two-beam interference patterns described in the main text, we occasionally observed fringe structures with more complex modulation. In a subset of shots, the spectral fringes exhibited additional structure beyond the standard two-beam pattern, qualitatively consistent with three-beam interference.
These patterns showed a primary fringe spacing similar to the two-beam case, but with extra modulation and intermediate maxima, suggesting the possible presence of more than two coherent emission contributions. While these features were not sufficiently frequent to support detailed quantitative analysis, their appearance indicates that under certain conditions the emission process may involve more complex interference behavior. Because these higher-order patterns were not the focus of the present study, they are noted here for completeness. The figure shows on the right, an example of single-shot data, and on the left, a simulation of 3-slit diffraction for comparison.